# $\mathcal{PT}$-symmetric time delay oscillator modelling beyond the weak coupling limit via a scattering matrix formulation

Abhijit Banerjee[1] & Trevor J. Hall[2]

[1]Academy of Technology, Department of Electronics & Communications Engineering, Adisaptagram, Hooghly, West Bengal, 712121, India.

[2]University of Ottawa, Electronic Engineering & Computer Science, Advanced Research Complex, 25 Templeton St, Ottawa, Ontario, K1N 6N5, Canada

## Abstract

Parity–time ($\mathcal{PT}$) symmetry in time-delay oscillators such as lasers and optoelectronic oscillators provides a potential route to enhanced spectral purity, including reduced phase noise and improved sidemode suppression. Existing theoretical descriptions are typically based on coupled-mode formulations derived under slowly varying envelope and near-degeneracy assumptions, which restrict their validity to weak coupling, small gain/loss contrast, and small detuning. In this work, a non-perturbative formulation of $\mathcal{PT}$-symmetric time-delay oscillators is developed based on a delay–difference equation and a scattering matrix representation of the coupling network. The approach treats propagation delay explicitly and does not rely on modal truncation, remaining valid for arbitrary coupling strength, gain/loss imbalance, and resonance detuning. The exact eigenvalue structure of the system is obtained in closed form, yielding a complete characterization of the unbroken and broken $\mathcal{PT}$-symmetric regimes as well as the associated exceptional points. A dimensionless order parameter is introduced that governs the symmetry transition over the full parameter space. It is further shown that conventional coupled-mode theory is recovered as an asymptotic limit of the exact formulation for small parameters. The results provide a unified and physically transparent framework for analysing $\mathcal{PT}$-symmetric delay systems beyond the weak-coupling limit, with direct implications for the design and optimisation of low-noise oscillators and photonic systems.

## 1 Introduction

Precision in time and frequency underpins advances in optical and wireless communications, radar, lidar, aerospace systems, and fundamental metrology. The performance of these systems is ultimately limited by fluctuations in their underlying oscillators. Time-delay oscillators (TDOs), including lasers [1] and optoelectronic oscillators (OEOs) [2], exploit delayed feedback to suppress noise and achieve high spectral purity. In such systems, large delays—realized, for example, by long optical cavities or fibreoptic delay lines—lead to dense mode spectra and strongly influence both stability and phase noise.

# $\mathcal{PT}$-symmetric time delay oscillator modelling beyond the weak coupling limit via a scattering matrix formulation

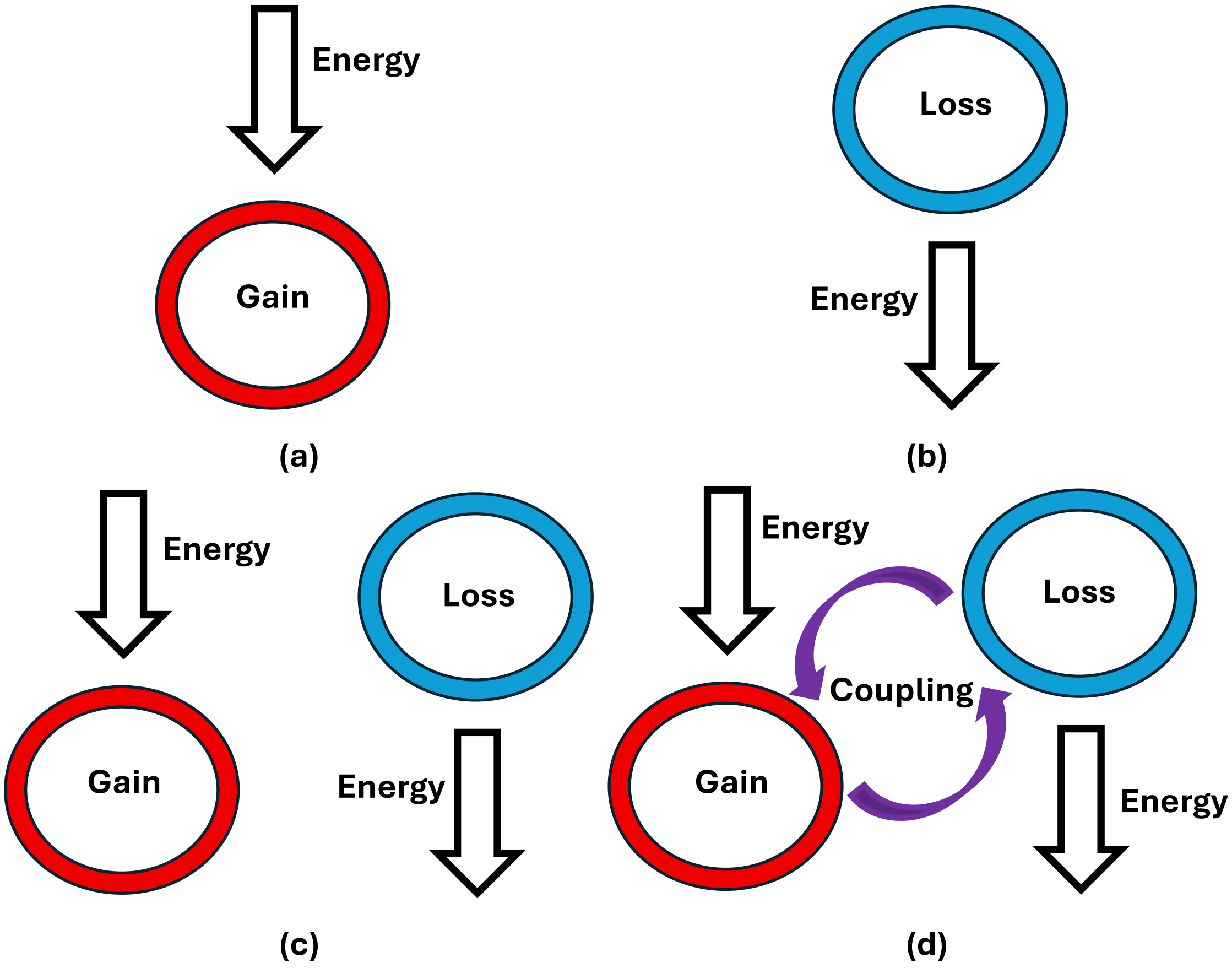

*Figure 1. Construction of an idealized $\mathcal{PT}$-symmetric oscillator system following [4]. (a) A subsystem in contact with the environment with net energy inflow providing gain. It cannot reach equilibrium. (b) A time reversed version of subsystem (a) with net energy outflow due to loss (i.e. dissipation). (c) A system composed of subsystems (a) & (b). The system is $\mathcal{PT}$-symmetric because under time reversal $\mathcal{T}$ loss and outflow of energy become gain and inflow of energy and under parity $\mathcal{P}$ the two subsystems are interchanged. However, in the absence of sufficient coupling, the system cannot achieve dynamic equilibrium, and the symmetry is effectively broken (d) A system composed of symmetrically coupled subsystems (a) & (b). If the net energy flows via the coupling from subsystem (a) flows rapidly enough into subsystem (b) then the system can be in dynamic equilibrium, and the $\mathcal{PT}$-symmetry is unbroken. Although both systems are in contact with the environment, the idealized system resembles an isolated system that conserves energy. This schematic is conceptual and does not represent a specific implementation; a concrete realisation in terms of coupled delay loops is introduced in Figure 2.*

Parity–time ($\mathcal{PT}$) symmetry, originally introduced by Bender and co-workers [3,4], provides a framework for understanding systems in which balanced gain and loss are arranged in a symmetric configuration. In its canonical form, a $\mathcal{PT}$-symmetric system is constructed by coupling a subsystem with its time-reversed counterpart such that parity exchanges the subsystems while time reversal interchanges gain and loss. Figure 1 presents the axiomatic construction of a $\mathcal{PT}$-symmetric system, in the context of classical oscillators, where energy flow replaces the probability flow of quantum systems, while Figure 2 introduces the specific realisation considered in this work. Depending on the relative strength of coupling and gain/loss, the system exhibits a phase transition between an unbroken regime, characterized by neutral mode doublets, and a broken regime, characterized by a growing and a decaying mode.

$\mathcal{PT}$-symmetry concepts have been explored *inter alia* in optical and laser physics [5-7] and optoelectronic oscillator systems [8,9] where coupled-loop or dual-path architectures exhibit phase-transition-like behavior associated with gain/loss balance and coupling parameters. While these implementations are commonly interpreted within a PT-symmetric framework, a rigorous realization of $\mathcal{PT}$-symmetry in time-delay oscillators requires a two-dimensional state space and a unimodular coupling operator, conditions that are not explicitly enforced in existing OEO architectures.

Coupled resonator systems exhibiting $\mathcal{PT}$-symmetry are commonly analyzed using a coupled-mode formalism [10] derived under slowly varying envelope and near-degeneracy assumptions. In these approaches, the fields of uncoupled resonators are used as basis functions, and coupling, detuning, and gain/loss are treated as perturbations. While this framework provides a convenient and widely used description, its validity is restricted to weak coupling, small detuning, and slowly varying dynamics. These assumptions can be problematic for time-delay oscillators, where the propagation delay is comparable to or larger than the characteristic timescales of the system and cannot be neglected.

In particular, in large-delay systems such as OEOs employing kilometre-scale fibre loops, the internal field cannot be regarded as spatially uniform, and the dynamics are inherently distributed in time. This motivates a formulation that treats delay explicitly, rather than as a perturbation.

In this work, a non-perturbative formulation of $\mathcal{PT}$-symmetric time-delay oscillators is developed based on a delay–difference equation and a scattering matrix description of the coupling network. The approach avoids modal truncation and does not rely on slowly varying envelope approximations, remaining valid for arbitrary coupling strength, gain/loss imbalance, and resonance detuning. The conditions for $\mathcal{PT}$-symmetry are expressed directly in terms of the transmission matrix of the coupling network, yielding a physically transparent parametrization in terms of coupling and gain/loss parameters.

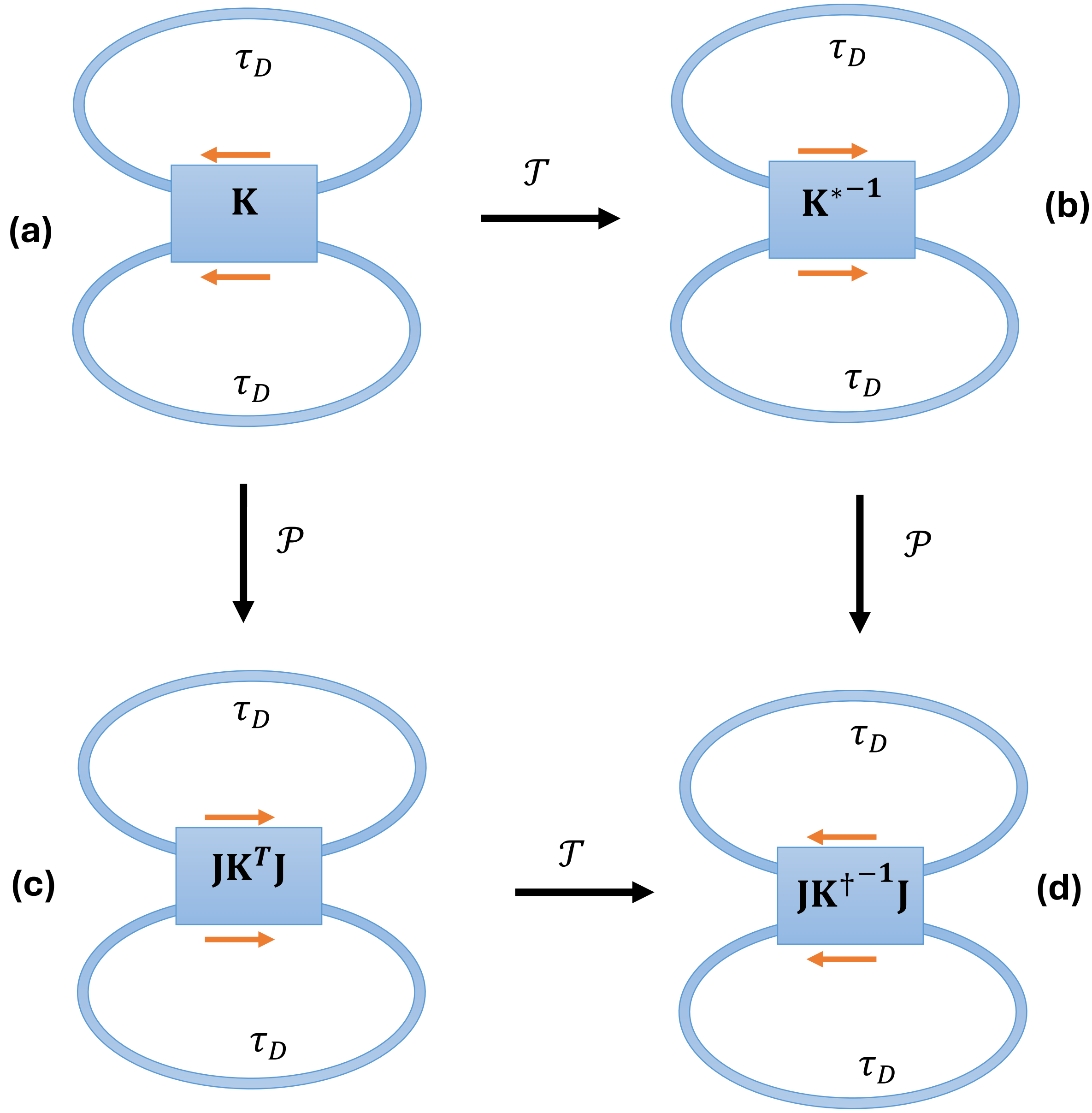

*Figure 2. An illustration of the commutation of the parity $\mathcal{P}$ and time $\mathcal{T}$ symmetry transformations applied to a coupled time delay oscillator. (a) The oscillator is composed of a pair of identical delay loops with delay time $\tau_D$ which are closed into a coupled pair of ring oscillators by the coupler represented by a rectangular block. The direction of propagation of the oscillations is shown by the red arrows. The coupler is characterised in the Fourier domain by a transmission matrix $\boldsymbol{K}$. (b) The transformation of the oscillator by time reversal: the propagation direction is reversed, gain and loss are interchanged but the delay loops are invariant. (c) The transformation of the oscillator by parity: the loops are swapped, the propagation direction is reversed, and the port labels exchanged. $\boldsymbol{J}$ is the exchange matrix (the first Pauli matrix) (d) The transformation of the oscillator by both parity and time reversal: the original propagation direction is restored.*

# $\mathcal{PT}$-symmetric time delay oscillator modelling beyond the weak coupling limit via a scattering matrix formulation

The exact eigenvalue structure of the system is derived, and the transition between unbroken and broken $\mathcal{PT}$-symmetric regimes is characterised in closed form. A dimensionless order parameter is introduced that controls this transition over the full parameter range. It is further shown that conventional coupled-mode theory emerges as an asymptotic limit of the exact formulation for small parameters, thereby establishing a direct connection between perturbative and non-perturbative descriptions.

The resulting framework provides a unified description of $\mathcal{PT}$-symmetric delay systems applicable to both weak and strong coupling regimes, with direct implications for the analysis and design of low-noise oscillators and photonic systems.

The paper is structured as follows. Section 2 introduces a delay-difference model of the coupled oscillator system and establishes the conditions for $\mathcal{PT}$-symmetry in terms of the coupler network transmission matrix. Section 3 derives the exact eigenvalue structure and characterises the unbroken and broken symmetry regimes. Section 4 shows that the conventional coupled-mode model arises as an asymptotic limit of the exact formulation. Section 5 summarises the main findings. Appendix I provides a detailed derivation of the scattering-matrix formulation and its reduction to the transmission model.

## 2 Delay difference equation model

Figure 2 illustrates a coupled pair of time-delay oscillators with identical roundtrip delay $\tau_D$.composed of a pair of lossless loops with identical delay time $\tau_D$ represented by an operator $\boldsymbol{\mathcal{D}}_{\tau_D}$ defined by:

$$\left(\boldsymbol{\mathcal{D}}_{\tau_D}\boldsymbol{u}\right)(t) = \boldsymbol{u}(t-\tau_D)$$

*Equation 1*

i.e., a time translation by the round-trip delay. The delay loops are interconnected by a linear time invariant network coupler represented by the operator $\boldsymbol{\mathcal{K}}$ so that:

$$\boldsymbol{u} = \kappa\boldsymbol{\mathcal{K}}\boldsymbol{\mathcal{D}}_{\tau_D}\boldsymbol{u}$$

*Equation 2*

where $\boldsymbol{u} \in \mathbb{C}^2$ is the oscillator state vector and $\kappa \in \mathbb{C}$ represents the common roundtrip gain and phase.

$\boldsymbol{\mathcal{K}}$ is characterised in the frequency domain by its transmission matrix $\mathbf{K}$. In a narrowband approximation around a carrier frequency $\omega_0$, the transmission $\mathbf{K}$ may be treated as frequency-independent and $\mathbf{K}(\omega_0)$ used directly in the time domain substituting for $\boldsymbol{\mathcal{K}}$**.** The frequency argument $\omega_0$ is suppressed for notational simplicity.

$\boldsymbol{\mathcal{D}}_{\tau_D}$ is also characterized in the frequency domain by a transmission matrix $\mathbf{D}_{\tau_D}$ given by:

$$\mathbf{D}_{\tau_D}(\omega) = \exp(-i\omega\tau_D)\,\mathbf{I}$$

*Equation 3*

The scalar $\kappa$ captures the net round-trip gain required for oscillation and the overall phase tuning of the loop, while the matrix $\mathbf{K}$ describes the redistribution of signals between the two delay loops.

## $\mathcal{PT}$-invariance

Appendix I introduces the action on a transmission matrix $\mathbf{T}$ of the parity $\mathcal{P}$ and time reversal $\mathcal{T}$ operators:

$$\mathbf{T} \overset{\mathcal{P}}{\rightarrow} \mathbf{J}\mathbf{T}^T\mathbf{J} \quad ; \quad \mathbf{T} \overset{\mathcal{T}}{\rightarrow} (\mathbf{T}^{-1})^* \quad ; \quad \mathbf{J} = \sigma_1 = \begin{bmatrix} 0 & 1 \\ 1 & 0 \end{bmatrix}$$

*Equation 4*

where the exchange matrix in two dimensions is represented by the first Pauli matrix $\sigma_1$.

The transmission matrix $\mathbf{D}_{\tau_D}$ is $\mathcal{P}$-invariant and $\mathcal{T}$-invariant. Consequently, the $\mathcal{PT}$-symmetry properties of the oscillator are determined by the coupler network which is $\mathcal{PT}$-invariant only if its transmission matrix $\mathbf{K}$ satisfies:

$$\mathbf{K} = \mathbf{J}(\mathbf{K}^{-1})^\dagger\mathbf{J} \quad \Leftrightarrow \quad \mathbf{K}\mathbf{J}\mathbf{K}^\dagger = \mathbf{J}$$

*Equation 5*

Equation 5 therefore imposes the condition that the coupling network preserve the pseudo-unitary metric associated with $\mathcal{PT}$-symmetry (see Appendix I).

Under this constraint, the transmission matrix takes the general form:

$$\mathbf{K} = \begin{bmatrix} a & ib \\ ic & d \end{bmatrix} \quad ; \quad |\det(\mathbf{K})| = 1 \quad ; \quad a, b, c, d \in \mathbb{R}$$

*Equation 6*

It is shown in Appendix I that a $\mathcal{PT}$-invariant transmission matrix admits the decomposition:

$$\mathbf{K} = \mathbf{U}(\beta)\boldsymbol{\Gamma}(\gamma)\mathbf{U}(\alpha) \quad ; \quad \alpha, \beta, \gamma \in \mathbb{R}$$

*Equation 7*

where, $\gamma$ parameterises the balanced gain/loss, while $\alpha$, $\beta$ are phase shift parameters that describe coupling-induced mixing between the two oscillator loops; $\mathbf{U}$ is a bisymmetric matrix describing a lossless coupler defined by:

$$\mathbf{U}(\theta) = \begin{bmatrix} \cos(\theta) & i\sin(\theta) \\ i\sin(\theta) & \cos(\theta) \end{bmatrix} \subset SU(2) \quad ; \quad \theta \in \mathbb{R}$$

*Equation 8*

with the useful property:

$$\mathbf{U}(\theta_1+\theta_2)=\mathbf{U}(\theta_1)\mathbf{U}(\theta_2)$$

*Equation 9*

and $\mathbf{\Gamma}$ is a real diagonal unit matrix describing balanced gain/ loss defined by:

$$\mathbf{\Gamma}(\gamma)=\begin{bmatrix}\exp(\gamma) & 0\\ 0 & \exp(-\gamma)\end{bmatrix}\subset SU(1,1)\quad ;\quad \gamma\in\mathbb{R}$$

*Equation 10*

Here $\gamma$ parameterizes the balanced gain–loss contrast between the two subsystems, while the parameters $\alpha$ and $\beta$ determine the effective coupling phase, with their sum $\alpha+\beta$ governing the eigenvalue structure (see Section 3).

Notably, $\mathbf{U}$ admits the decomposition:

$$\mathbf{U}(\theta)=\mathbf{H}\exp(i\theta\sigma_3)\,\mathbf{H}\quad ;\quad \mathbf{H}=\frac{1}{\sqrt{2}}\begin{bmatrix}1 & 1\\ 1 & -1\end{bmatrix}\quad ;\quad \sigma_3=\begin{bmatrix}1 & 0\\ 0 & -1\end{bmatrix}$$

*Equation 11*

which describes the transmission of a Mach-Zehnder interferometer with a differential phase shift of $\theta$ in its arms represented by $\exp(i\theta\sigma_3)$ formed between two 180° hybrid couplers represented by $\mathbf{H}$; a real Hadamard matrix in Sylvester's form normalised to be orthogonal. $\sigma_3$ is the third Pauli matrix.

Equivalent representations of $\mathbf{K}$ related by similarity transformations yield identical eigenvalue structure, emphasising that only the combined parameters governing the pseudo-unitary coupling affect the modal behaviour.

# 3 Eigenvalue structure

The eigenvalue structure of the transmission matrix $\mathbf{K}$ determines the modal behaviour of the coupled oscillator system. In particular, the interplay between coupling and gain/loss governs whether the system operates in the unbroken or broken $\mathcal{PT}$-symmetric regime. This transition is reflected in the nature of the eigenvalues, which either lie on the unit circle or form a reciprocal real pair.

The substitutions:

$$\boldsymbol{u}=\boldsymbol{a}\exp(st)\quad ;\quad s=\sigma+i\omega$$

*Equation 12*

$$\kappa\exp(-s\tau_D)\,\lambda=1$$

*Equation 13*

where $\boldsymbol{a} \in \mathbb{C}^2$ is a constant coefficient, transforms Equation 2 into the algebraic eigenvalue-eigenvector problem:

$$\mathbf{K}\boldsymbol{a} = \lambda \boldsymbol{a}$$

*Equation 14*

which has characteristic polynomial:

$$\lambda^2 - \mathrm{tr}(\mathbf{K})\,\lambda + \det(\mathbf{K}) = 0$$

*Equation 15*

Equation 13 is a statement of the Barkhausen oscillation condition.

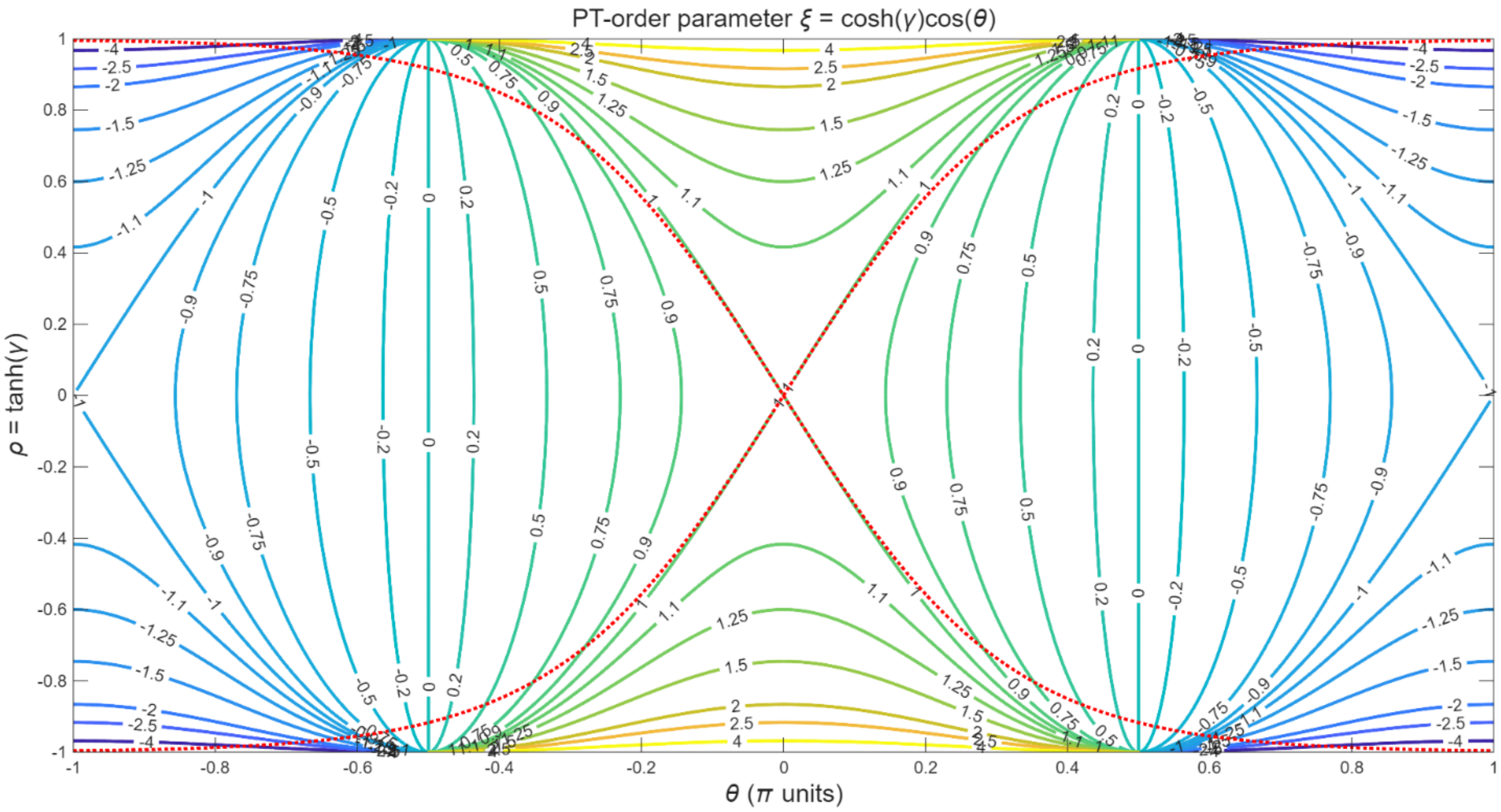

*Figure 3. The level curves of the $\mathcal{PT}$-order parameter $\xi$ over its entire domain. The exceptional point contours with $\xi^2 = 1$ delineate regions with unbroken $\mathcal{PT}$-symmetry ($\xi^2 < 1$) and broken $\mathcal{PT}$-symmetry ($\xi^2 > 1$). The exceptional point contours according to the coupled mode formulism are shown by the dotted red curves.*

It has been assumed that the coupler network is sufficiently broadband that it may be treated as frequency independent. Nevertheless, an overall bandwidth limitation about some nominal operating frequency may be modelled by a single pole low pass-filter acting on the complex envelope of the oscillation. This converts the delay difference equation to a delay differential equation. In the frequency domain this change introduces a frequency dependent common gain:

$$\kappa(s) = \kappa(0)\frac{1}{1 + is\tau_R}$$

*Equation 16*

into the Barkhausen oscillation condition. The eigen structure of $\mathbf{K}$ is undisturbed. The time constant $\tau_R$ is the pass-band centre group delay introduced by the band limiting component. In what follows, for simplicity of exposition, the description is limited to the broadband case, which is accurate for the large number of modes that fall well within the passband of a time delay oscillator with large delay $\tau_D \gg \tau_R$.

The characteristic equation (Equation 15) has *real* coefficients on account of Equation 6. Since the trace is equal to the sum of the eigenvalues and the determinant is equal to their product, the eigenvalues $\lambda$ either form a complex conjugate pair corresponding to solutions with *unbroken $\mathcal{PT}$-symmetry* or a reciprocal pair on the real axis corresponding to solutions with *broken $\mathcal{PT}$-symmetry*.

While the extension to complex frequencies $s$ provides valid solutions for all $\sigma$, for non-zero $\sigma$ the delay operator Equation 3 breaks time reversal symmetry, thus strictly unbroken $\mathcal{PT}$-symmetry modes must be neutral, i.e. neither grow nor decay which requires $|\kappa| = 1$. In practice, the gain control mechanism of an oscillator, typically amplifier saturation, maintains this condition for equilibrium solutions.

Explicitly, the eigenvalues are given by:

$$\lambda_{\pm} = (\mathrm{tr}(\mathbf{K})/2) \pm \sqrt{(\mathrm{tr}(\mathbf{K})/2)^2 - \det(\mathbf{K})}$$

*Equation 17*

Making use of the one parameter subgroup property (Equation 9) of $\mathbf{U}$, the invariance of the trace and determinant to cyclic permutations, and the unimodularity condition $\det(\mathbf{K}) = 1$, Equation 17 simplifies to:

$$\lambda_{\pm} = \xi \pm \sqrt{\xi^2 - 1}$$

*Equation 18*

where a $\mathcal{PT}$-order parameter $\xi$ has been introduced:

$$\xi(\theta, \gamma) = \cosh(\gamma)\cos(\theta) \quad ; \quad \theta = \alpha + \beta$$

*Equation 19*

The parameter $\xi$ thus provides a compact measure of the relative strength of coupling and gain/loss in the system.

Notably, the symmetry transition determined by the order parameter $\xi$ is independent of the common loop gain $\kappa$ and therefore is not influenced by frequency-dependent amplification within the oscillator bandwidth.

Since the eigenvalue structure depends only on the sum $\theta = \alpha + \beta$, the two coupling stages are not independently observable at the level of modal dynamics and may be reduced to a single equivalent coupling parameter.

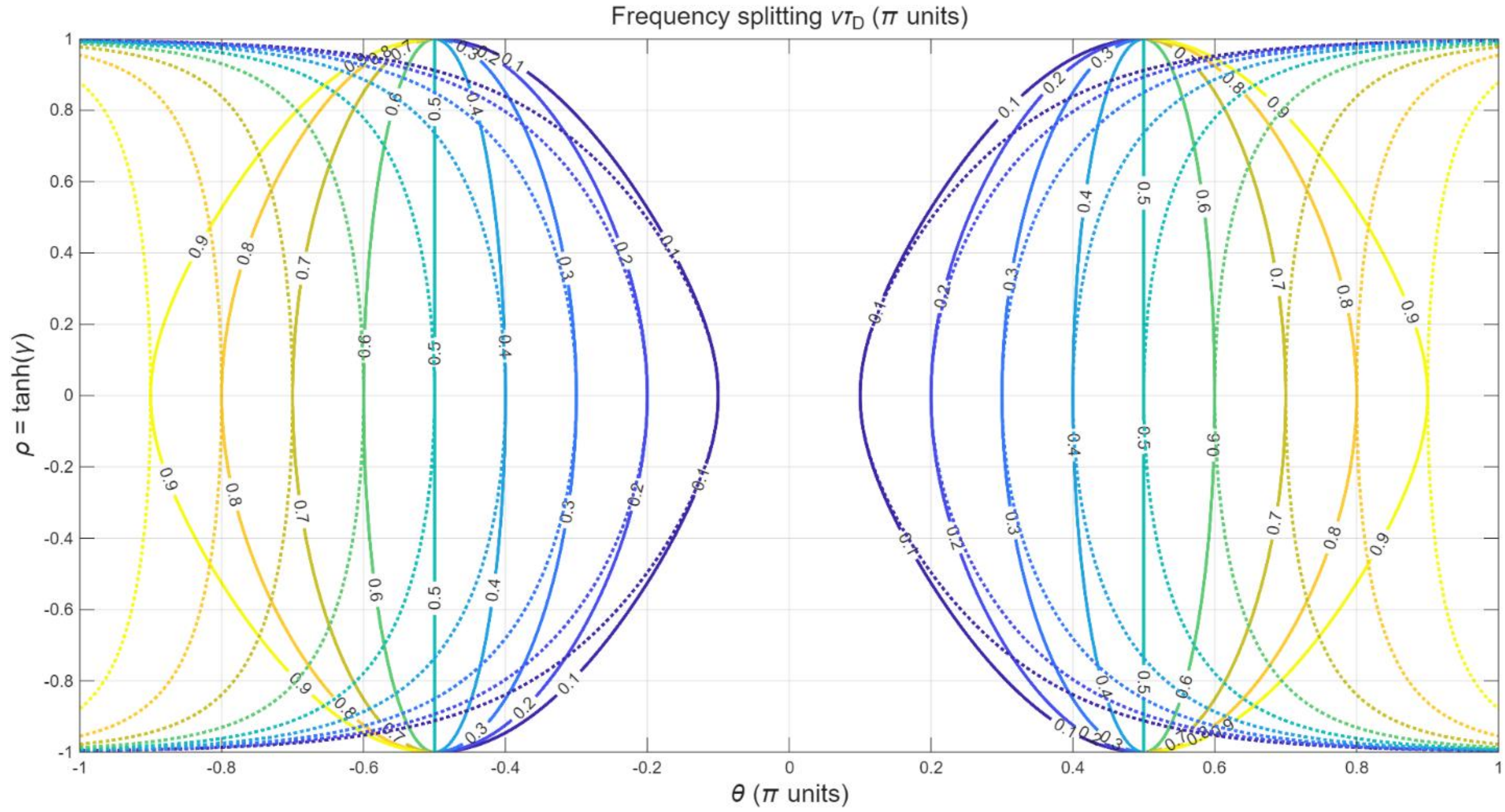

*Figure 4. The level curves of the frequency splitting parameter $\nu\tau_D$ over its entire domain. The level curves of $\nu\tau_D$ according to the coupled mode formulism are shown by the dotted curves.*

Noting that the $\mathcal{PT}$-order parameter is periodic in $\theta$ and that the subsidiary parameter:

$$\rho = \tanh(\gamma) \in [-1,1]$$

*Equation 20*

is monotonic in $\gamma$ and has finite range, the infinite domain of the $\mathcal{PT}$-order parameter $\xi$ may be mapped onto a finite domain in the variables $\theta$ and $\rho$.

Figure 3 shows that the exceptional point contours are determined solely by the coupling and gain–loss parameters and are independent of the overall loop gain. This indicates that the symmetry transition is a global property of the coupled system rather than a mode-selective effect.

## Unbroken $\mathcal{PT}$-symmetry regime

The discriminant in Equation 17 is negative for:

$$\xi^2 < 1$$

*Equation 21*

which yields a complex conjugate pair of eigenvalues on the unit circle with frequency splitting ν given by:

$$\lambda_{\pm} = \exp(\pm i\nu\tau_D)$$

*Equation 22*

$$\nu\tau_D = \tan^{-1}\left(\frac{\sqrt{1-\xi^2}}{\xi}\right)$$

*Equation 23*

The arctangent in Equation 23 may be interpreted in the four-quadrant sense to account for a change in sign of $\xi$. The level curves of $\nu\tau_D$ are plotted in Figure 4.

The resonances occur in doublets with each resonance of the doublet displaced in frequency by $\pm\nu$ from the nodes of a regular grid with interval $2\pi/\tau_D$. The maximum doublet splitting is $\pi/\tau_D$ corresponding to a regular grid with interval $\pi/\tau_D$. An alternative perspective is that the resonances correspond to two interlaced regular frequency combs with interval $2\pi/\tau_D$ that are displaced respectively by $\pm\nu$ from complete alignment.

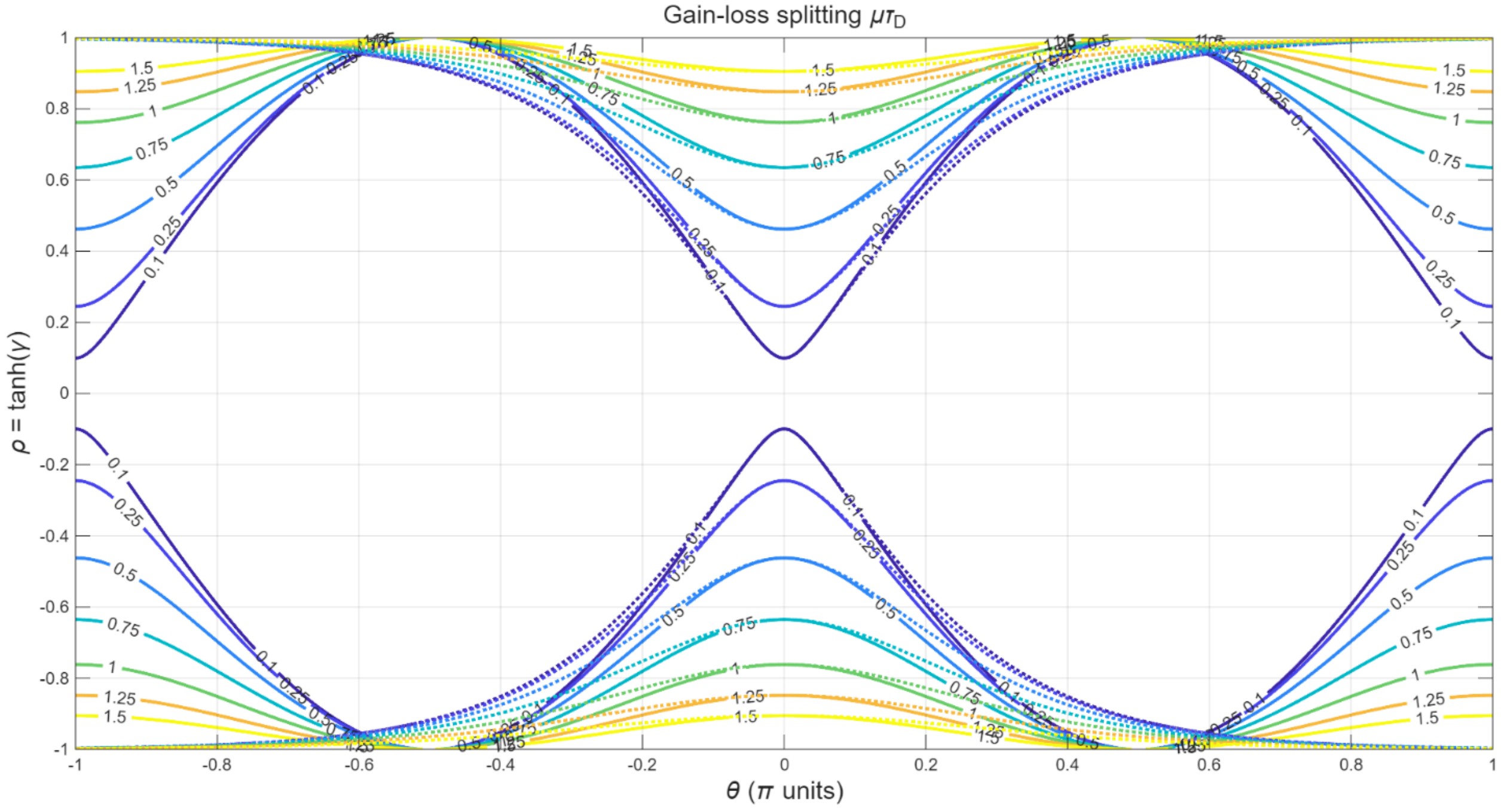

*Figure 5. The level curves of the gain-loss splitting parameter $\mu\tau_D$ over its entire domain. The level curves of $\mu\tau_D$ according to the coupled mode formulism are shown by the dotted curves.*

## Broken $\mathcal{PT}$-symmetry regime

The discriminant in Equation 17 is positive for:

$$\xi^2 > 1$$

*Equation 24*

which leads to a reciprocal pair of eigenvalues on the positive ($\xi > 1$) or negative ($\xi < -1$) real axis with gain-loss splitting $\mu$ given by:

$$\lambda_{\pm} = \mathrm{sgn}(\xi)\exp(\pm\mu\tau_D) \quad ; \quad \mu\tau_D = \ln\left(|\xi| + \sqrt{\xi^2 - 1}\right)$$

*Equation 25*

For $|\kappa| = 1$ the two modes exhibit reciprocal amplification and attenuation, with one growing and the other decaying in time.

The resonant frequencies form a regular grid with interval $2\pi/\tau_D$.

## The exceptional points

The unbroken and broken $\mathcal{PT}$-symmetry regimes are delineated by a zero discriminant that occurs for:

$$\xi^2 = 1$$

*Equation 26*

The contour $\xi^2(\theta, \gamma) = 1$ on the coupling-gain plane $(\theta, \gamma)$ delineates the regions of unbroken and broken $\mathcal{PT}$-symmetry (see Figure 3). Points on this contour are known as exceptional points. As an exceptional point is approached the two eigenvalues approach each other to form in the limit a unit magnitude double real root, i.e., the algebraic multiplicity is 2. The two eigenvectors associated with the two eigenvalues also approach each other. In the limit the matrix $\mathbf{K}$ becomes non-diagonalizable (defective), possessing a single eigenvector.

# 4 Coupled mode model

The $\mathcal{PT}$-symmetric coupler network transmission matrix $\kappa\mathbf{K}$ for small $\theta$, $\gamma$, $\ln(\kappa)$, may be expanded as a power series which up to first order is given by:

$$\kappa\mathbf{K} = \kappa\mathbf{U}(\theta)\mathbf{\Gamma}(\gamma) = \mathbf{I} + \ln(\kappa)\,\mathbf{I} + \begin{bmatrix} 0 & i\theta \\ i\theta & 0 \end{bmatrix} + \begin{bmatrix} \gamma & 0 \\ 0 & -\gamma \end{bmatrix} + \cdots$$

*Equation 27*

Hence Equation 2 becomes:

$$\boldsymbol{u}(t) - \boldsymbol{u}(t - \tau_D) = \left(\begin{bmatrix} ln(\kappa) + \gamma & i\theta \\ i\theta & ln(\kappa) - \gamma \end{bmatrix} + \cdots\right)\boldsymbol{u}(t - \tau_D)$$

*Equation 28*

At zeroth order:

$$\boldsymbol{u}(t) - \boldsymbol{u}(t - \tau_D) = \mathbf{0}$$

*Equation 29*

which has an arbitrary $\tau_D$-periodic function as solution which may be represented by a Fourier series:

$$\boldsymbol{u}(t) = \sum_p \boldsymbol{a_p} \exp(i\omega_p t) \quad ; \quad \omega_p \tau_D = 2p\pi \quad p \in \mathbb{Z}$$

*Equation 30*

That is there are a multitude of super-modes each with a *constant* complex amplitude $\boldsymbol{a_p} \in \mathbb{C}^2$.

For small perturbations the complex amplitudes $\boldsymbol{a_p}$ will become slowly varying functions of time. Substitution of Equation 30 into Equation 28 then yields:

$$\boldsymbol{a}(t) - \boldsymbol{a}(t-\tau_D) = \left(\begin{bmatrix} \ln(\kappa) + \gamma & i\theta \\ i\theta & \ln(\kappa) - \gamma \end{bmatrix} + \cdots \right) \boldsymbol{a}(t-\tau_D)$$

*Equation 31*

where $\boldsymbol{a}$ is any one of the $\boldsymbol{a_p}$.

A Taylor series expansion:

$$\boldsymbol{a}(t-\tau_D) = \boldsymbol{a}(t) - \tau_D \frac{d\boldsymbol{a}}{dt}(t) + \cdots$$

*Equation 32*

leads up to first order accuracy to the differential system:

$$\tau_D \frac{d\boldsymbol{a}}{dt} = \begin{bmatrix} \ln(\kappa) + \gamma & i\theta \\ i\theta & \ln(\kappa) - \gamma \end{bmatrix} \boldsymbol{a}$$

*Equation 33*

This equation is the standard coupled-mode description and is recovered here as the leading-order approximation of the exact delay formulation.

Equation 33 admits solutions of the form:

$$\boldsymbol{a}(t) = \boldsymbol{c} \exp(st)$$

*Equation 34*

where $s\tau_D$ is the eigenvalue and $\boldsymbol{c}$ is its associate eigenvector solutions of the eigenvalue-eigenvector problem:

$$\begin{bmatrix} \ln(\kappa) + \gamma & i\theta \\ i\theta & \ln(\kappa) - \gamma \end{bmatrix} \boldsymbol{c} = s\tau_D \boldsymbol{c}$$

*Equation 35*

## $\mathcal{PT}$-symmetric time delay oscillator modelling beyond the weak coupling limit via a scattering matrix formulation

The coupled mode and the delay-difference equations provide a solution for the oscillator state vector of the same form but while the eigenvalues $s\tau_D$ of the delay difference equation model satisfy

$$s\tau_D = \ln(\kappa) + \ln\left(\cos(\theta)\cosh(\gamma) \pm \sqrt{(\cos(\theta)\cosh(\gamma))^2 - 1}\right)$$

*Equation 36*

the coupled mode model eigenvalues $s\tau_D$ satisfy:

$$s\tau_D = \ln(\kappa) \pm \sqrt{\gamma^2 - \theta^2}$$

*Equation 37*

A straightforward calculation confirms the asymptotic agreement of Equation 36 and Equation 37 for small $\gamma$, $\theta$. The level curves of $\nu\tau_D$ and $\mu\tau_D$ according to the coupled mode formulism illustrated by the dotted contours in Figure 4 and Figure 5 respectively, confirm this asymptotic agreement.

# 5 Conclusions

A non-perturbative formulation of $\mathcal{PT}$-symmetric time-delay oscillators has been developed based on a delay–difference equation combined with a scattering matrix description of the coupling network. By explicitly incorporating propagation delay, the formulation avoids the slowly varying envelope and modal truncation assumptions that underlie conventional coupled-mode approaches.

The condition for $\mathcal{PT}$-symmetry has been expressed directly in terms of the transmission matrix of the coupling network, leading to a physically meaningful decomposition into coupling and gain/loss components. This enabled the exact eigenvalue structure of the system to be derived in closed form, providing a complete characterization of the unbroken and broken symmetry regimes, as well as the associated exceptional points, across the full parameter space.

A dimensionless order parameter has been introduced to describe the symmetry transition, yielding compact expressions for both frequency splitting in the unbroken regime and gain–loss splitting in the broken regime. These results extend the analysis of $\mathcal{PT}$-symmetric systems to regimes of strong coupling and large delay that are not accessible within perturbative frameworks.

It has been further demonstrated that conventional coupled-mode theory arises naturally as an asymptotic limit of the present formulation for small gain/loss and coupling parameters. This establishes a direct connection between perturbative and non-perturbative descriptions, clarifying the range of validity of widely used models.

The framework developed here provides a rigorous and unified basis for analysing $\mathcal{PT}$-symmetric delay systems and opens the way to systematic exploration of large-delay and strong-coupling effects in practical implementations. Extensions to include dispersion, bidirectional propagation, and nonlinear saturation mechanisms will enable direct application to the prediction and optimisation of phase noise and mode selection in advanced oscillator architectures. In particular, the identification of a global symmetry transition independent of the overall loop gain provides a basis for distinguishing intrinsic symmetry effects from frequency-selective behaviour arising from external filtering or gain dispersion in time-delay oscillators.

# References

[1] M. Alouini, G. Danion, M. Vallet, ‘Self-linewidth-narrowing photonic oscillator’, Opt. Exp. **33**, 1021-1032, (2025).

[2] F. Zou, L. Zou, B. Yang, Q. Ma, X. Zou, J. Zou, S. Chen, D. Milosevic, Z. Cao, H. Liu3, ‘Optoelectronic oscillator for 5G wireless networks and beyond’, J. Phys. D, **54**:423002, (2021).

[3] C. M. Bender, S. Boettcher, ‘Real spectra in non-Hermitian Hamiltonians having PT symmetry’, Phys. Rev. Lett., **80**(24), 5243-5246, (1998).

[4] C. M. Bender, ‘PT symmetry: in quantum and classical physics’, (World Scientific, 2019).

[5] R. El-Ganainy, K. G. Makris, M. Khajavikhan, Z. H. Musslimani, Stefan Rotter, D. N. Christodoulides, ‘Non-Hermitian physics and PT symmetry’, Nature Physics, **14** (1), 11-19, (2018).

[6] H. Zhao, L. Feng, ‘Parity–time symmetric photonics’, National Science Review **5**: 183–199, (2018).

[7] B. Qi, H.-Z. Chen, L. Ge, P. Berini, R.-M. Ma, ‘Parity–time symmetry synthetic lasers: physics and devices." Advanced Optical Materials. **7**(22),1900694 (2019).

[8] Y. Liu, T. Hao, W. Li, J. Capmany, N. Zhu, M. Li ‘Observation of parity-time symmetry in microwave photonics’, Light: Science & Applications 2018 ;**7**(1):38 (2018).

[9] C. Huan, Z. Li, E.H.W. Chan, Y. Li, P. Hao, X. Wang, X. S. Yao, ‘Parity time symmetric optoelectronic oscillator with low phase noise and high sidemode suppression ratio’, J. Lightwave Technology, **43**(14), 6613-6619, (2025).

[10] H. Haus, ‘Waves and fields in optoelectronics’, (Prentice Hall, 1984).

# $\mathcal{PT}$-symmetric time delay oscillator modelling beyond the weak coupling limit via a scattering matrix formulation

## Appendix I Scattering matrices & $\mathcal{PT}$-symmetry

### Scattering matrix description

A linear time-invariant system with $n$ ports is characterised in the *frequency domain* by its scattering matrix $\mathbf{S} \in \mathbb{C}^{n \times n}$ which maps incoming mode amplitudes $\mathbf{a} \in \mathbb{C}^n$ to outgoing mode amplitudes $\mathbf{b} \in \mathbb{C}^n$ by:

$$\mathbf{b} = \mathbf{S}\mathbf{a}$$

*Equation 38*

In a narrowband approximation around a carrier frequency $\omega_0$, the scattering matrix may be treated as frequency-independent and $\mathbf{S}(\omega_0)$ used directly in the time domain. For notational simplicity it is then common to suppress its argument $\omega_0$.

### $\mathcal{PT}$-symmetry

Time reversal $\mathcal{T}: t \mapsto -t$ maps incoming to outgoing modes via complex conjugation and inversion of the scattering process:

$$\mathbf{S} \overset{\mathcal{T}}{\rightarrow} (\mathbf{S}^{-1})^*$$

*Equation 39*

Consequently:

- time reversal acts *anti-unitarily*;
- unitary matrices are transposed;
- propagation delays are invariant;
- gain and loss are exchanged;

Parity $\mathcal{P}: \boldsymbol{x} \mapsto -\boldsymbol{x}$ reverses the propagation direction and exchanges port labels. Its action on a scattering matrix is a *persymmetric* transformation:

$$\mathcal{P}: \mathbf{S} \mapsto \mathbf{J}\mathbf{S}^{\mathrm{T}}\mathbf{J}$$

*Equation 40*

where $\mathbf{J}$ is the *exchange matrix* that has all trailing diagonal entries set to unity and all entries elsewhere set to zero.

This operation:

- exchanges ingress and egress ports (encoded by the conjugation by $\mathbf{J}$)
- reverses propagation direction (encoded by the transposition)
- swaps gain and loss subsystems

The actions of $\mathcal{P}$ and $\mathcal{T}$ on scattering matrices inherit the properties of $\mathcal{P}$,$\mathcal{T}$ as commuting involutions: Their combined action on the scattering matrix is:

$$\mathbf{S} \xrightarrow{\mathcal{PT}} \mathbf{J}(\mathbf{S}^{-1})^{\dagger}\mathbf{J}$$

*Equation 41*

Thereby a $\mathcal{PT}$-symmetric system satisfies:

$$\mathbf{S} = \mathbf{J}(\mathbf{S}^{-1})^{\dagger}\mathbf{J} \quad \Leftrightarrow \quad \mathbf{S}\mathbf{J}\mathbf{S}^{\dagger} = \mathbf{J}$$

*Equation 42*

Although gain/loss and spatial exchange individually break the symmetry, the combined $\mathcal{PT}$ operation restores invariance. Notably taking the determinant of the $\mathcal{PT}$-symmetry condition implies that the determinant of the scattering matrix is unimodular.

$$\det(\mathbf{S})\det(\mathbf{J})\det(\mathbf{S}^{\dagger}) = \det(\mathbf{J}) \quad \Longrightarrow \quad |\det(\mathbf{S})|^2 = 1$$

*Equation 43*

## $\mathcal{PT}$-symmetric scattering matrix decomposition

Specialising to a 4-port network and denoting the exchange matrix in $n$ dimensions by $\mathbf{J}_n$, the exchange matrix $\mathbf{J}_4$ admits the partition:

$$\mathbf{J}_4 = \begin{bmatrix} \mathbf{0} & \mathbf{J}_2 \\ \mathbf{J}_2 & \mathbf{0} \end{bmatrix}$$

*Equation 44*

Introduce a change of basis transformation matrix:

$$\mathbf{P} = \frac{1}{\sqrt{2}}\begin{bmatrix} \mathbf{I_2} & \mathbf{J}_2 \\ \mathbf{J}_2 & -\mathbf{I_2} \end{bmatrix}$$

*Equation 45*

where $\mathbf{I_2}$ is the identity in two dimensions.

$\mathbf{P}$ is a real, symmetric, unitary and hence involutory matrix:

$$\mathbf{P}^{\dagger}\mathbf{P} = \mathbf{I} \quad \Longrightarrow \quad \mathbf{P}^2 = \mathbf{I}$$

*Equation 46*

that block-diagonalizes $\mathbf{J}_4$:

$$\mathbf{P}\mathbf{J}_4\mathbf{P}^{-1} = \begin{bmatrix} \mathbf{I_2} & \mathbf{0} \\ \mathbf{0} & -\mathbf{I_2} \end{bmatrix} = \boldsymbol{\eta}$$

*Equation 47*

into signature form $\boldsymbol{\eta} = \mathrm{diag}(\mathbf{I_2}, -\mathbf{I_2})$.

Introducing the representation:

$$\mathbf{S} = \mathbf{P}\mathbf{W}\mathbf{P}^{-1} = \mathbf{P}\mathbf{W}\mathbf{P}$$

*Equation 48*

then:

$$\mathbf{S}\mathbf{J}_4\mathbf{S}^\dagger = \mathbf{J}_4 \quad \Leftrightarrow \quad \mathbf{W}\boldsymbol{\eta}\,\mathbf{W}^\dagger = \boldsymbol{\eta}$$

*Equation 49*

That is $\mathbf{W}$ is an element of the pseudo-unitary group of transformations $\mathrm{U}(2,2)$ that preserve the indefinite metric defined by $\boldsymbol{\eta}$.

$\mathbf{W}$ admits the Cartan (KAK) decomposition of $\mathrm{U}(2,2)$:

$$\mathbf{W} = \mathbf{U_1}\boldsymbol{\Sigma}\mathbf{U_2}$$

*Equation 50*

where $\mathbf{U_1}, \mathbf{U_2}$ take the form:

$$\mathbf{U} = \begin{bmatrix} \mathbf{U}_+ & \mathbf{0} \\ \mathbf{0} & \mathbf{U}_- \end{bmatrix} \quad ; \quad \mathbf{U}_+, \mathbf{U}_- \in \mathrm{U}(2)$$

*Equation 51*

and $\boldsymbol{\Sigma}$ takes the form:

$$\boldsymbol{\Sigma} = \begin{bmatrix} \cosh(\boldsymbol{\gamma}) & \sinh(\boldsymbol{\gamma}) \\ \sinh(\boldsymbol{\gamma}) & \cosh(\boldsymbol{\gamma}) \end{bmatrix} \quad ; \quad \boldsymbol{\gamma} = \begin{bmatrix} \gamma_1 & 0 \\ 0 & \gamma_2 \end{bmatrix}$$

*Equation 52*

where $\mathbf{U}_+, \mathbf{U}_-$ carry the compact degrees of freedom and $\boldsymbol{\Sigma}$ carries the non-compact degrees of freedom $(\boldsymbol{\gamma})$.

## Reduced transmission model

Under the assumption of negligible backscattering and unidirectional propagation[1] the $\mathcal{PT}$-symmetry condition reduces to:

$$\mathbf{K} = \mathbf{J_2}(\mathbf{K^{-1}})^\dagger\mathbf{J_2} \quad \Leftrightarrow \quad \mathbf{K}\mathbf{J_2}\mathbf{K}^\dagger = \mathbf{J_2}$$

*Equation 53*

where $\mathbf{K}$ is the transmission matrix component of the scattering matrix $\mathbf{S}$:

$$\mathbf{K} = \begin{bmatrix} s_{31} & s_{32} \\ s_{41} & s_{42} \end{bmatrix}$$

*Equation 54*

In what follows the dimension of the matrices is clear from the context consequently for simplicity dimensional subscripts are now dropped.

The reduction of dimension reduces the change of basis transformation to a $2 \times 2$ Hadamard matrix:

[1] This assumption corresponds physically to directional coupling or isolation. It simplifies analysis but excludes bidirectional coupling and mode hybridisation.

$$\mathbf{H} = \frac{1}{\sqrt{2}}\begin{bmatrix} 1 & 1 \\ 1 & -1 \end{bmatrix}$$

*Equation 55*

that diagonalizes the exchange matrix into signature form $\boldsymbol{\eta} = \mathrm{diag}(1, -1)$.

Introducing the representation:

$$\mathbf{K} = \mathbf{H}\mathbf{W}\mathbf{H}^{-1} = \mathbf{H}\mathbf{W}\mathbf{H}$$

*Equation 56*

Then:

$$\mathbf{S}\mathbf{J}\mathbf{S}^{\dagger} = \mathbf{J} \quad \Leftrightarrow \quad \mathbf{W}\boldsymbol{\eta}\,\mathbf{W}^{\dagger} = \boldsymbol{\eta}$$

*Equation 57*

That is $\mathbf{W}$ is an element of the pseudo-unitary group of transformations $\mathrm{U}(1,1)$ that preserve the indefinite metric defined by $\boldsymbol{\eta}$.

$\mathbf{W}$ admits the Cartan (KAK) decomposition of $\mathrm{U}(1,1)$:

$$\mathbf{W} = \mathbf{U_1}\boldsymbol{\Sigma}\mathbf{U_2}$$

*Equation 58*

where $\mathbf{U_1}, \mathbf{U_2}$ take the form:

$$\mathbf{U} = \begin{bmatrix} \mathrm{U}_+ & 0 \\ 0 & \mathrm{U}_- \end{bmatrix} \quad ; \quad \mathrm{U}_+, \mathrm{U}_- \in \mathrm{U}(1)$$

*Equation 59*

and $\boldsymbol{\Sigma}$ takes the form:

$$\boldsymbol{\Sigma} = \begin{bmatrix} \cosh(\gamma) & \sinh(\gamma) \\ \sinh(\gamma) & \cosh(\gamma) \end{bmatrix} \in \mathrm{SU}(1,1)$$

*Equation 60*

where $\mathrm{U}_+, \mathrm{U}_-$ carry the compact degrees of freedom and $\boldsymbol{\Sigma}$ carries the non-compact degree of freedom $(\gamma)$.

In the two-dimensional case it is convenient to absorb any overall phase shift within the common tuning phase and chose $\mathbf{U_1}, \mathbf{U_2}$ to take the form:

$$\exp(i\theta\sigma_3) = \begin{bmatrix} \exp(i\theta) & 0 \\ 0 & \exp(-i\theta) \end{bmatrix} \quad ; \quad \sigma_3 = \begin{bmatrix} 1 & 0 \\ 0 & -1 \end{bmatrix}$$

*Equation 61*

where $\sigma_3$ is the third Pauli matrix.

This leads to the representation:

$$\mathbf{K} = \mathbf{U}(\beta)\boldsymbol{\Gamma}(\gamma)\mathbf{U}(\alpha)$$

*Equation 62*

where:

$$\mathbf{\Gamma_\gamma} = \exp(\gamma\sigma_3) = \begin{bmatrix} \exp(\gamma) & 0 \\ 0 & \exp(-\gamma) \end{bmatrix}$$

*Equation 63*

represents balanced gain and loss and:

$$\mathbf{U}(\theta) = \mathbf{H}\exp(i\theta\sigma_3)\,\mathbf{H} \equiv \begin{bmatrix} \cos(\theta) & i\sin(\theta) \\ i\sin(\theta) & \cos(\theta) \end{bmatrix}$$

*Equation 64*

represents a lossless coupler with a bisymmetric transmission which may be interpreted physically as the transmission matrix of a Mach-Zehnder interferometer (MZI) with a differential phase shift of $\theta$ its arms formed between two 50:50 180° hybrid couplers represented by the Hadamard matrices $\mathbf{H}$.

$\mathrm{SU}(1,1)$ describes hyperbolic rotations preserving an indefinite metric, corresponding physically to power exchange between gain and loss channels.

## Scope and limitations

The $\mathrm{SU}(1,1)$ structure in the prequel relies on:

- negligible backscattering;
- effective unidirectional propagation.

In the general bidirectional case:

- $\mathcal{PT}$-symmetry still requires $\mathbf{SJS}^\dagger = \mathbf{J}$ but no reduction to a $2 \times 2$ matrix exists;
- the system belongs instead to a higher-dimensional pseudo-unitary group.

The $\mathrm{SU}(1,1)$ structure is exact for the reduced model but not for the full scattering problem.

## Summary

- $\mathcal{PT}$-symmetry equips the scattering matrix with pseudo-unitary structure reducing to $SU(1,1)$ under unidirectional propagation.
- This admits a physically meaningful decomposition:

  coupler⟶gain/loss⟶coupler
- Relaxing the no-backscattering assumption yields a more general $U(2,2)$ system with richer dynamics but without $SU(1,1)$ reduction.